\documentclass[twocolumn,showpacs,preprintnumbers,amsmath,amssymb,superscriptaddress]{revtex4}

\usepackage{graphicx}
\begin{document}
\bibliographystyle{prsty}
\title{Electronic excitations of magnetic impurity state in diluted magnetic semiconductor (Ga,Mn)As
}

\author{M.~Kobayashi}
\affiliation{Department of Applied Chemistry, 
School of Engineering, The University of Tokyo, 
7-3-1 Hongo, Bunkyo-ku, Tokyo 113-8656, Japan}
\affiliation{Synchrotron Radiation Research Organization, 
University of Tokyo, 1-490-2 Kouto, Sayo-cho, Tatsuno, 
Hyogo 679-5165, Japan}
\author{H.~Niwa}
\affiliation{Department of Applied Chemistry, 
School of Engineering, The University of Tokyo, 
7-3-1 Hongo, Bunkyo-ku, Tokyo 113-8656, Japan}
\author{Y.~Takeda}
\affiliation{Synchrotron Radiation Research Unit, 
Japan Atomic Energy Agency, Sayo-gun, Hyogo 679-5148, Japan}
\author{A.~Fujimori}
\affiliation{Department of Physics, The University of Tokyo, 
7-3-1 Hongo, Bunkyo-ku, Tokyo 113-0033, Japan}
\author{Y.~Senba}
\affiliation{Japan Synchrotron Radiation Research Institute (JASRI), Sayo, Hyogo 679-5198, Japan}
\author{H. Ohashi}
\affiliation{Japan Synchrotron Radiation Research Institute (JASRI), Sayo, Hyogo 679-5198, Japan}
\author{A. Tanaka}
\affiliation{Department of Quantum Matter, ADSM, Hiroshima University, Higashi-Hiroshima 739-8530, Japan}
\author{S.~Ohya}
\affiliation{Department of Electronic Engineering, The University of Tokyo, 
Hongo, Bunkyo-ku, Tokyo 113-8656, Japan}
\author{P.~N.~Hai}
\affiliation{Department of Electronic Engineering, The University of Tokyo, 
Hongo, Bunkyo-ku, Tokyo 113-8656, Japan}
\author{M.~Tanaka}
\affiliation{Department of Electronic Engineering, The University of Tokyo, 
Hongo, Bunkyo-ku, Tokyo 113-8656, Japan}
\author{Y.~Harada}
\affiliation{Synchrotron Radiation Research Organization, 
University of Tokyo, 1-490-2 Kouto, Sayo-cho, Tatsuno, 
Hyogo 679-5165, Japan}
\affiliation{Institute for Solid State Physics, The University of Tokyo, 
1-1-1 Koto, Sayo, Hyogo 679-5198, Japan}
\author{M.~Oshima}
\affiliation{Department of Applied Chemistry, 
School of Engineering, The University of Tokyo, 
7-3-1 Hongo, Bunkyo-ku, Tokyo 113-8656, Japan}
\affiliation{Synchrotron Radiation Research Organization, 
University of Tokyo, 1-490-2 Kouto, Sayo-cho, Tatsuno, 
Hyogo 679-5165, Japan}
\date{\today}

\begin{abstract}
The electronic structure of doped Mn in (Ga,Mn)As is studied by resonant inelastic X-ray scattering (RIXS). 
From configuration-interaction cluster-model calculations, the line shapes of the Mn $L_3$ RIXS spectra can be explained by $d$-$d$ excitations from the Mn$^{3+}$ ground state, dominated by charge-transferred states, rather than a Mn$^{2+}$ ground state. 
Unlike archetypical $d$-$d$ excitation, the peak widths are broader than the experimental energy resolution. 
We attribute the broadening to a finite lifetime of the $d$-$d$ excitations, which decay rapidly to electron-hole pairs in the host valence and conduction bands through hybridization of the Mn $3d$ orbital with the ligand band. 
\end{abstract}

%\pacs{71.28.+d, 71.30.+h, 79.60.Dp, 73.61.-r}

\maketitle

%\section{Introduction}
A diluted magnetic semiconductor (DMS), in which a host semiconductor is doped with a low concentrations of magnetic ions, is a material of interest for spintronics research. In particular, ferromagnetic DMSs have attracted considerable attention because the itinerant carriers are considered to mediate the magnetic interaction between doped ions \cite{NatPhys_10_Sawicki, Science_11_Yamada}. 
This type of magnetism is called ``carrier-induced ferromagnetism'' and enables the possibility of manipulating both electronic charge and spin degrees of freedom. 
III-V based DMS Ga$_{1-x}$Mn$_x$As is an archetypical ferromagnetic DMS and has been studied from both fundamentally and in applications as a ferromagnetic DMS \cite{RMP_06_Jungwirth}. Spintronic devices using Ga$_{1-x}$Mn$_x$As have been fabricated and their effectiveness has been demonstrated \cite{Nature_00_Ohno, Nature_08_Chiba}, although the Curie temperature ($T_{\mathrm{C}}$) of Ga$_{1-x}$Mn$_x$As is below room temperature ($T_{\mathrm{C}} < 200$ K).

Understanding the mechanism of ferromagnetism in Ga$_{1-x}$Mn$_x$As is strongly desirable to develop DMS for application in spintronic devices. 
Several physical models of the ferromagnetism have been proposed, e.g., the Zener $p$-$d$ exchange model \cite{Science_00_Dietl, PRB_07_Jungwirth, NatPhys_10_Sawicki}, the Mn $3d$ impurity band model \cite{PRL_01_Berciu, PRL_06_Burch, NatPhys_11_Ohya}, and the magnetic polaron model \cite{PRL_02_Kaminski, PRB_02_Mayr}. These models depend on the localization of hole carriers around the Mn ions, the energy position of the Mn $3d$ states, and the strength of hybridization between the Mn $3d$ orbital and the ligand band. 
To obtain a fundamental understanding of ferromagnetism, it is essential to know the Mn $3d$ electronic configurations in the ground state. If the valence state of Mn in the ground state is divalent (Mn$^{2+}$), the Mn ion acts as an acceptor supplying a hole to the host GaAs, supporting the Zener $p$-$d$ exchange model. 
Thus, the electronic structure of Mn $3d$ ions in Ga$_{1-x}$Mn$_x$As has been intensively investigated experimentally \cite{NatPhys_11_Ohya, PRB_98_Okabayashi, PRB_01_Okabayashi, PRL_11_Edmonds}. However, the presence of divalent (Mn$^{2+}$) or trivalent (Mn$^{3+}$) states has not been conclusively determined, and the degree of the localization of the Mn $3d$ states remains a matter of dispute.

In this Letter, to address the electronic structure of the doped Mn ions, we report the results of Mn $L_3$ X-ray absorption spectroscopy (XAS) and resonant inelastic X-ray scattering (RIXS) measurements of Ga$_{1-x}$Mn$_x$As ($x=0.04$). 
RIXS is a powerful tool to investigate electronic excitations in element- and symmetry-specific ways including $d$-$d$ and charge-transfer (CT) excitations for open shell $3d$ orbitals \cite{PRB_06_Ghiringhelli}, and magnetic excitations for spin/charge-ordered systems \cite{NatPhys_11_LeTacon}. 
These excitations are sensitive to electron correlation, crystalline symmetry, and the strength of hybridization with ligand band. 
Moreover, RIXS is less sensitive to the surface structure because of its long probing depth ($>100$ nm). 
The RIXS spectra obtained for Ga$_{1-x}$Mn$_x$As are compared with configuration-interaction (CI) cluster-model calculations \cite{JSPS_94_Tanaka}, and the electronic structure parameters are estimated. 
It is found that the bulk electronic structures of Mn ions in Ga$_{1-x}$Mn$_x$As consist predominantly of charge-transferred states. 
Based on the experimental findings, we shall discuss the Mn $3d$ electronic state in Ga$_{1-x}$Mn$_x$As.

%\section{Experimental}
Ga$_{1-x}$Mn$_x$As ($x=0.04$) thin films with a thickness of 20 nm were grown on a GaAs(001) substrate at 270 $^{\circ}$C under ultra-high vacuum by a molecular beam epitaxy method. To avoid surface oxidation, the samples were covered by GaAs (2 nm) and an As capping layer (1 nm) after the deposition of Ga$_{1-x}$Mn$_x$As layer to produce a structure of As/GaAs/Ga$_{0.96}$Mn$_{0.04}$As/GaAs(buffer)/GaAs(001). 
The thin film sample after the deposition was cut into two pieces and one half of the sample was annealed at 270 $^{\circ}$C for 2 hours. 
$T_\mathrm{C}$ values of the as-grown and annealed samples were $\sim 65$ K and $\sim 100$ K, respectively, as determined by the Arrott plot of the magnetic circular dichroism. 
The RIXS and XAS experiments were performed using the high-resolution soft X-ray emission station HORNET \cite{RSI_12_Harada} at the long undulator beamline BL07LSU of SPring-8, Japan \cite{NIMPRESA_11_Senba}. The total energy resolution for the RIXS experiments was about 170 meV at the Mn $L_3$ edge ($E/\Delta E > 3750$). The RIXS spectra were measured with linear horizontal polarization at room temperature under a vacuum of $1.0 \times 10^{-5}$ Pa. 
The XAS signals were collected in the total-fluorescence-yield (TFY) mode. The XAS spectra of MnO were measured in the total-electron-yield (TEY) mode as a reference.

\begin{figure}[t!]
\begin{center}
\includegraphics[width=8.8cm]{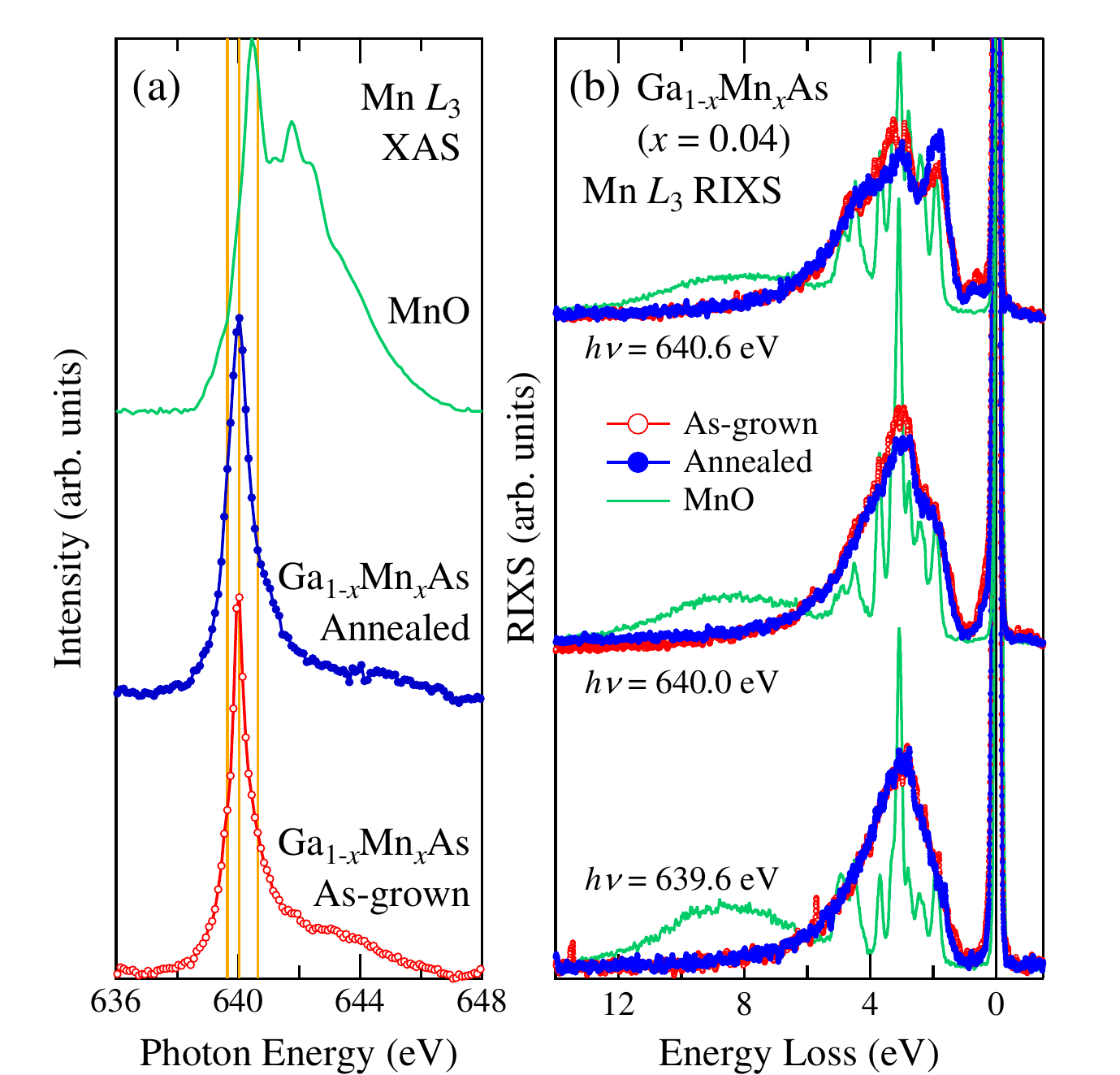}
\caption{Mn $L_{3}$ resonant inelastic X-ray scattering spectra of the as-grown and annealed Ga$_{1-x}$Mn$_x$As ($x = 0.04$) thin films. 
(a) Mn $L_3$ XAS spectra taken in the TFY mode. 
The vertical lines denote the excitation photon energy for the RIXS measurements. 
(b) Mn $L_{3}$ RIXS spectra. The excitation energies are 639.6, 640.0, and 640.6 eV. 
The spectra of MnO are also shown as a reference. 
}
\label{GMA_RIXS}
\end{center}
\end{figure}

\begin{figure}[t!]
\begin{center}
\includegraphics[width=8.8cm]{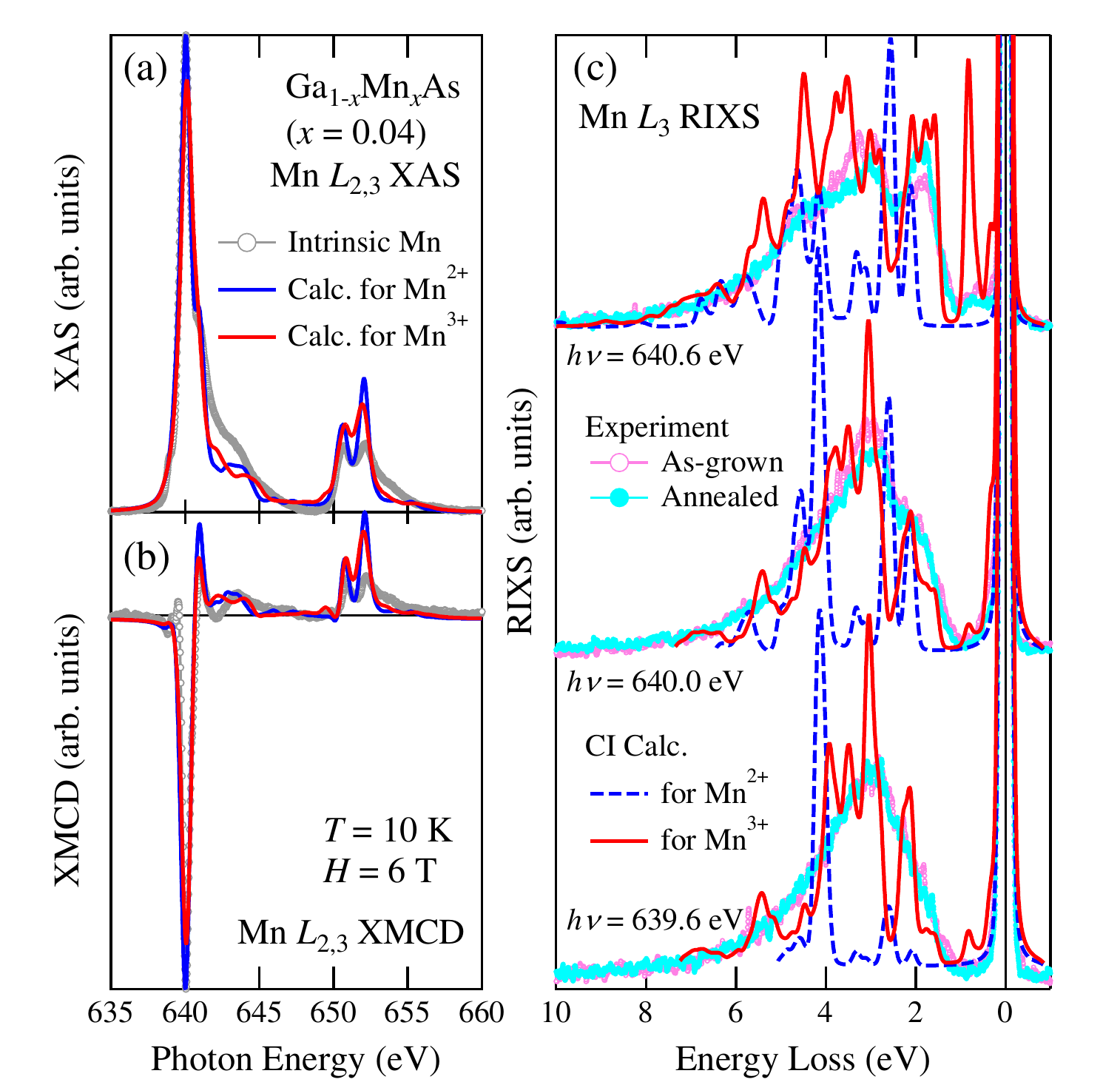}
\caption{RIXS spectra compared with the configuration-interaction (CI) cluster-model calculations. 
(a), (b) Comparison of the intrinsic Mn $L_{2,3}$ XAS and XMCD spectra with the CI calculations, respectively. The experimental XAS and XMCD spectra are extracted from the magnetic-field dependence of XMCD \cite{PRL_08_Takeda}. 
(c) Comparison of experimental RIXS spectra with CI calculations for Mn$^{2+}$ and Mn$^{3+}$ states. The CI calculations for RIXS were performed using the parameters common to those of XAS (XMCD). The calculated spectra are broadened by the experimental energy resolution of 170 meV. 
}
\label{GMA_CIcalc}
\end{center}
\end{figure}

%\section{Results and discussion}
%Mn L23 XAS
Figure~\ref{GMA_RIXS}(a) shows the Mn $L_3$ XAS spectra of the Ga$_{1-x}$Mn$_x$As thin films. 
The XAS spectra show a broad profile compared with the reference MnO, similar to the XAS specta previously reported for chemically etched Ga$_{1-x}$Mn$_x$As thin films \cite{APL_04_Edmonds, PRB_05_Edmonds}. 
The XAS spectra exhibit the same features as those of the ``intrinsic Mn component'' [see Fig.~\ref{GMA_CIcalc}(a)], which consists mainly of Mn ions substituting Ga sites and partial filling of some interstitial site \cite{PRL_08_Takeda}. 
In Ga$_{1-x}$Mn$_x$As, there are ferromagnetic and paramagnetic Mn components, which come from the intrinsic Mn and extrinsic oxidized Mn ions, respectively. 
Details for the assignment of Mn species are described in Ref.~\cite{PRL_08_Takeda}. 
It has been reported that low-temperature annealing of Ga$_{1-x}$Mn$_x$As induces interstitial Mn defects to diffuse toward the surface \cite{PRL_04_Edmonds, NatMater_12_Dobrowolska} creating Mn oxides at the surface \cite{PRB_05_Ishiwata}. 
However, because of bulk sensitivity of the fluorescence yield mode, the XAS spectra do not show the oxide components and reflects the bulk electronic structure.

%Mn L3 RIXS
Figure~\ref{GMA_RIXS}(b) shows the Mn $L_3$ RIXS spectra of the Ga$_{1-x}$Mn$_x$As thin films taken with various excitation (incident photon) energies $h\nu$ across the $L_3$ edge in the XAS spectra. The spectra are normalized by area and plotted as a function of energy loss ($E_L$). 
The line shape of the RIXS spectra strongly depends on the excitation energy. As a reference, Mn $L_3$ RIXS spectra of MnO for the corresponding excitation energy are also shown in Fig.~\ref{GMA_RIXS}(b). 
The RIXS spectra of MnO demonstrate sharp $d$-$d$ excitation peaks from $E_L = 2 - 6$ eV and broad features taht aoriginate from CT excitation above $E_L = 6$ eV \cite{PRB_06_Ghiringhelli}. Notably, the energy resolution is high enough to resolve individual $d$-$d$ excitation peaks in MnO. 
In contrast, the RIXS spectra of Ga$_{1-x}$Mn$_x$As show broad line shapes even at the same high energy resolution, which are similar to those previously reported at lower energy resolution \cite{PRB_05_Ishiwata}. 
Therefore, the broad features of the Mn $L_3$ RIXS spectra are expected to accurately reflect the electronic structure of the Mn ions in Ga$_{1-x}$Mn$_x$As. 
Since the CT excitation features from MnO are absent, the CT excitation energy is likely to be reduced to an energy level comparable to that of the $d$-$d$ excitation ($E_L = 2 - 6$ eV) and contribute to the broadening of the $d$-$d$ transition peaks \cite{PRB_06_Taguchi}. 
There are differences in the RIXS spectra taken at $h\nu = 640$ and 640.6 eV between the as-grown and annealed samples. 
These changes in the RIXS spectra taken at $h\nu = 640$ and 640.6 eV by annealing may be attributed to out-diffusion of interstitial Mn atoms during annealing.

%XAS vs Calc
The intrinsic Mn $L_{2,3}$ XAS and X-ray magnetic circular dichroism (XMCD) spectra are compared with CI cluster-model calculations, as shown in Fig.~\ref{GMA_CIcalc}(a) and \ref{GMA_CIcalc}(b), respectively. The calculations are performed for Mn ions in divalent (Mn$^{2+}$) or trivalent (Mn$^{3+}$) states in a crystal field of tetrahedral ($T_d$) symmetry. 
Both the calculated spectra for the Mn$^{2+}$ and Mn$^{3+}$ states reproduce the experimental spectra. Common electronic structure parameters are used to fit bot the XAS and XMCD: the ligand-to-$3d$ CT energy $ \Delta = 1.5$ eV, the $d$-$d$ Coulomb interaction energy $U_{dd} = 3.5$ eV, the Slater-Koster parameter $(pd\sigma) = -1.0$ eV, and the crystal-field splitting $10Dq = 0$ eV. 
Here, the CT energy $\Delta$ is defined as the energy difference between the $d^5$ and $d^6\underline{L}$ states for Mn$^{2+}$, and between $d^5\underline{L}$ and $d^6\underline{L}^2$ states for Mn$^{3+}$, where $\underline{L}$ denotes a hole in the ligand bands. 
The values of $\Delta$, $U_{dd}$, and $(pd\sigma)$ are the same as those estimated from the photoemission experiments \cite{PRB_99_Okabayashi}. 
We note that spectra calculated using $10Dq$ larger than 0.3 eV could not reproduce the experimental spectra. 
The small crystal-field splitting is similar to that of the iron pnictide BaFe$_2$As$_2$ \cite{FPC_10_Ma} and the chalcopyrite CuFeS$_2$ \cite{pssa_09_Sato}, as expected from the small electronegativity of the heavy chalcogenide anions compared with that of oxygen \cite{FN_CF}.

%RIXS vs Calc
The CI calculations for the Mn$^{2+}$ and Mn$^{3+}$ states were also performed for the RIXS spectra using the same electronic structure parameters as those for XAS and XMCD. 
Figure~\ref{GMA_CIcalc}(b) shows the Mn $L_3$ RIXS spectra of the Ga$_{1-x}$Mn$_x$As thin film compared with the CI calculations broadened by a Gaussian of the experimental energy resolution. The calculated spectra for the ``Mn$^{3+}$'' state reproduce the energy distribution of the experimental spectra except for the linewidths of each $d$-$d$ excitation peak, while the calculated spectra for the Mn$^{2+}$ state are different from the experimental RIXS spectra. 
It should be noted that the profiles of the RIXS spectrum for the annealed film are also close to the calculated spectrum for the Mn$^{3+}$ state. Since annealing drives the interstitial Mn atoms towards the surface \cite{PRB_05_Edmonds} and increases the number of Mn in substitutional sites relative to the interstitial site \cite{NatMater_12_Dobrowolska}, the spectral change induced by the annealing may be attributed to the relative increase of Mn atoms in substituted sites. 
Interstitial Mn defects are expected to act as double donors and have similar electronic structure to the CI calculations for Mn$^{2+}$. The relative decrease of Mn atoms in interstitial sites induced by annealing may also cause the peak at $E_L = 2$ eV to be more pronounced. 
The results suggest that the doped Mn atoms in bulk Ga$_{1-x}$Mn$_x$As are predominantly in the formally trivalent Mn$^{3+}$ state. In addition, the electronic configurations in the Mn$^{3+}$ ground state are calculated as $3d^4: 3d^5\underline{L}: 3d^6\underline{L}^2: 3d^7\underline{L}^3 = 9\%: 69\% : 21\% : 1\%$. 
Here, it is clearly shown that the delocalized charge-transferred states ($3d^5\underline{L}$ and $3d^6\underline{L}^2$) are dominant in the electronic configurations of the Mn$^{3+}$ ground state, and are inherently different from the ``Mn$^{2+}$'' ground state with a localized $d^5$ state hybridized with $d^6\underline{L}$ state. 
Thus the Mn $L_3$ RIXS spectra are very sensitive to the valence state of Mn and our results conclusively show ``Mn$^{3+}$'' in the ground state of Ga$_{1-x}$Mn$_x$As.

%Discussion
Based on our experimental findings, we discuss the detail of the Mn $3d$ electronic states in Ga$_{1-x}$Mn$_x$As. 
The transport properties of Ga$_{1-x}$Mn$_x$As have been reported to show a metal-insulator transition at $T_\mathrm{C}$ with metallic behavior below $T_\mathrm{C}$ \cite{PRB_98_Matsukura}, implying that the hole carriers are not itinerant above $T_\mathrm{C}$. 
This transport property has been qualitatively well explained by bound magnetic polarons, in which Mn ions weakly bind holes \cite{PRB_03_Kaminski}. 
Because the occupied Mn $3d$ levels are located below the As $4p$ ligand band \cite{PRB_99_Okabayashi} and the present results suggest that the hybridization between the $3d$ orbital and the As $4p$ ligand is strong enough, the Anderson impurity model predicts the formation of a split-off state above the valence-band maximum \cite{PRB_01_Okabayashi}. 
The split-off state acts as an acceptor level and leads to the formation of the impurity band \cite{PRB_01_Okabayashi}, in which the Mn ion weakly binds a hole, which may be the origin of the bound magnetic polarons. 
Our RIXS results show that the dominant charge transferred states are the trivalent Mn$^{3+}$ ground state and indicates the presence of a ligand hole bound to the Mn configuration. 
This arrangement will significantly reduce ionized impurity scattering and support the presence of quantum states in Ga$_{1-x}$Mn$_x$As quantum wells observed by resonant tunneling spectroscopy \cite{NatPhys_11_Ohya, PRL_10_Ohya}.

\begin{figure}[t!]
\begin{center}
\includegraphics[width=6cm]{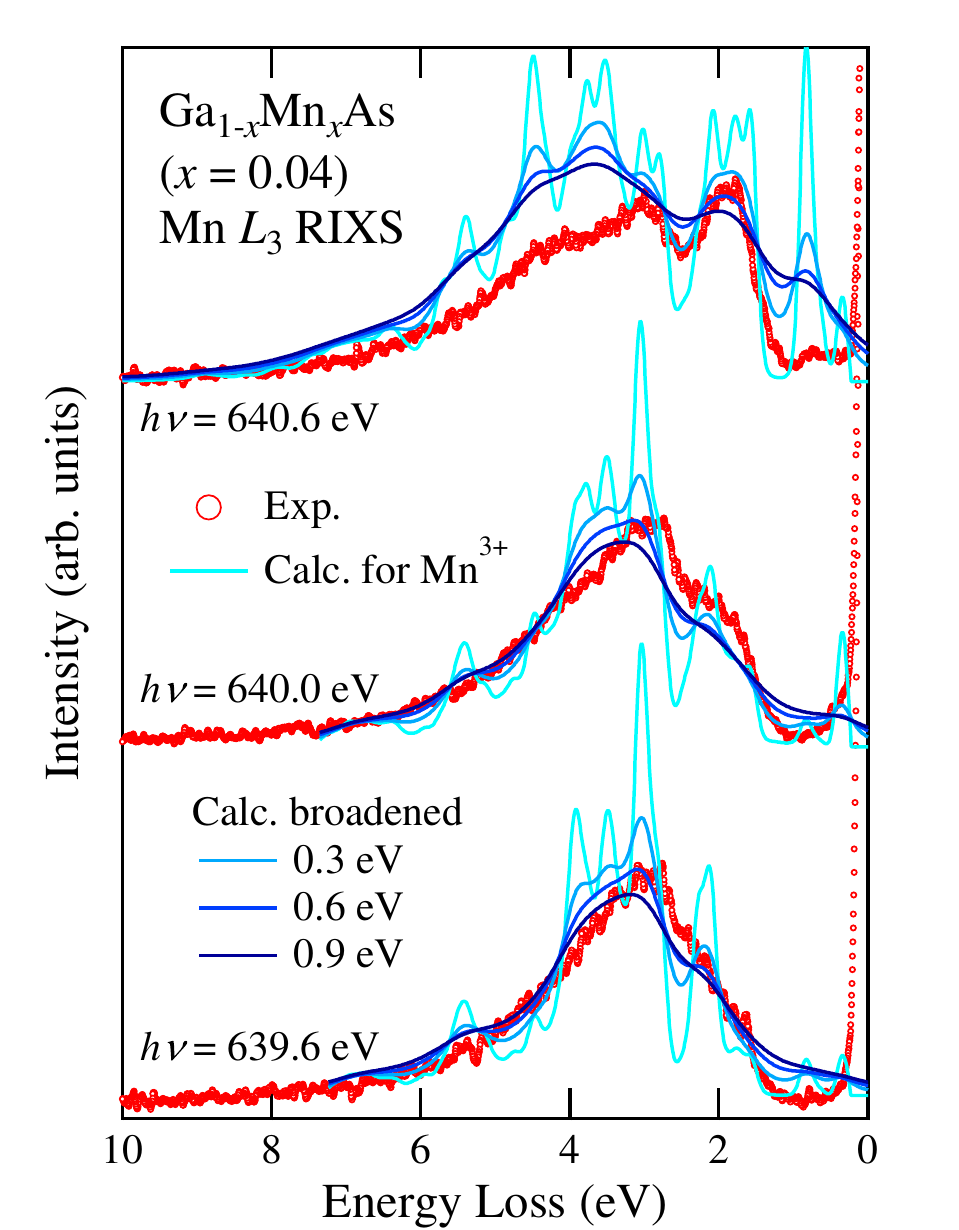}
\caption{Lorentzian broadening of the RIXS spectra calculated by the CI cluster model. 
The experimental RIXS spectra of the annealed sample are compared with the calculations for the Mn$^{3+}$ configuration convolved by Lorentzian lineshapes. The full-width-half-maximum values of the Lorentzian the broadening are 0.3, 0.6, and 0.9 eV. 
}
\label{GMA_LB}
\end{center}
\end{figure}

The RIXS spectra are now well reproduced by the CI cluster-model calculation using a common parameter set with XAS and XMCD. However, the experimental RIXS profile is still broader than the calculated profile with 0.17 eV broadening because of the experimental resolution. 
To evaluate the additional broadening of the Mn $d$-$d$ excitation, the CI calculations are further broadened. 
The calculated RIXS spectra with additional broadening are shown in Fig.~\ref{GMA_LB} and broadened spectra with a full-width-half-maximum of 0.6 eV reproduce the experimental RIXS spectra relatively well. 
According to Ref.~\cite{PRB_06_Taguchi}, the RIXS spectra can be broadened because of stronger hybridization between $d$-$d$ excitations and particle-hole pairs in the host valence and conduction bands itself. 
However, this effect should be accompanied by distortion of the overall RIXS profile \cite{PRB_06_Taguchi} and may fail to explain either the experimental RIXS or XAS/XMCD spectra. 
Another reason could be a double exchange interaction between the $d$ orbitals \cite{PR_51b_Zener} but this hardly contributes to the broadening of the Mn $3d$ states because hole carriers have $p$-type ligand character as discussed above. 
Alternatively, we can attribute the additional broadening to a lifetime broadening in the RIXS final state where fast decay of the $d$-$d$ excitation into an electron-hole pair in the host valence and conduction bands occurs through the hybridization between the Mn $3d$ state and the As $4p$ ligand band. 
To account for this, we have applied Lorentzian, rather than the Gaussian, broadening to the RIXS spectra as shown in Fig.~\ref{GMA_LB}. 
The additional broadening implies two steps for the RIXS decay process; First, the resonantly excited core-hole state decays into a $d$-$d$ excitation involving a localized bound hole state (split-off state) because of the strong hybridization between the Mn $3d$ and As $4p$ orbitals. Then, through the interaction with the As $4p$ band, the $d$-$d$ excitation decays into the electron-hole pair in the host valence and conduction bands. 
With increasing energy resolution, RIXS may provide information about the time evolution of the decay process, and expand the applications of RIXS in the near future.

%\section{Conclusion}
In conclusion, we have conducted Mn $L_3$ RIXS and XAS measurements on Ga$_{1-x}$Mn$_x$As ($x=0.04$) thin films to investigate the electronic structure of the doped Mn ions. 
The XAS spectrum taken in the TFY mode includes few contributions from extrinsic surface Mn oxides, suggesting that the TFY-XAS spectra largely reflect the bulk electronic structure of Ga$_{1-x}$Mn$_x$As. 
The RIXS spectra, which also reflect bulk sensitive information, show a broad profile even though at energy resolutions high enough to distinguish individual $d$-$d$ excitation peaks of MnO. 
Analysis by the CI calculations indicates that the Mn ground states mainly consists of the Mn$^{3+}$ electronic configuration composed of the charge-transferred states ($d^5\underline{L}$ and $d^6\underline{L}^2$), in which the ligand hole is weakly bound to the Mn $3d^5$ state, rather than the pure Mn$^{2+}$ state. 
The width of the Mn $d$-$d$ excitation has been estimated to be $\sim 0.6$ eV by the Lorentzian broadening of the calculated spectra. 
Therefore, the broad $d$-$d$ excitation peaks in the RIXS spectra can be attributed to the lifetime broadening in the final state of the RIXS process where fast decay of the $d$-$d$ excitations to an electron-hole pair in the host valence and conduction bands occurs because of the hybridization between the Mn $3d$ orbital and the ligand band.

%\section{Acknowledgments}
This work was carried out by a joint research in the Synchrotron Radiation Research Organization, The University of Tokyo (Proposal No. 2011A7403 and 2009BS03). 
M.K. acknowledges support from the Japan Society for the Promotion of Science.


\begin{thebibliography}{10}
\bibitem{FN_CF}
The crystal field includes the ionic and the covalent contributions. When the
  hybridization between $d$ orbital and the ligand band is not negligible, the
  ionic contribution becomes small or negative due to expansion of the charge
  distribution. The covalent contribution is basically positive.
\end{thebibliography}
\end{document}